# Towards an Abelian Formulation of Lattice QCD Confinement

Ken Yee

Dept. of Physics and Astronomy, L.S.U.

Baton Rouge, Louisiana   70803-4001

email: kyee@rouge.phys.lsu.edu

### Abstract

We probe for operators occurring in the APQCD("abelian-projected QCD") action by evaluating abelian-projected 1-plaquette spectral densities in pure gauge $SU(3)$ fixed to maximal abelian gauge. Couplings $\overline{B}_{APQCD}(q,L)$ are extracted from the spectral densities for each representation $q$, $L \times L$ plaquette. While APQCD is dominated by a $q = L = 1$ resonance, we also find evidence for weakly coupled $L = 2$ plaquettes. Moreover, since $\overline{B}_{APQCD}(1,1) > \overline{B}_{QED}(1,1)$ even if $\beta_{QED} > \beta_c$, $L > 1$ plaquettes must be significant since APQCD is confining.

# 1 Lattice Action for APQCD

As in gauge-Higgs systems [1], magnetic monopole singularities can exist in pure gauge QCD [2]. The significance of these monopoles, as exemplified in compact QED [3], lies in their possible responsibility for QCD confinement. In the lattice abelian-projection approach to finding them [4], $SU(N)$ gauge symmetry is fixed to maximal abelian(MA) gauge, which has residual $[U(1)]^{N-1}$ gauge symmetry. Performing an abelian projection yields a $[U(1)]^{N-1}$ lattice gauge theory with $N$ constrained abelian gauge fields("species") corresponding to the $N$ diagonal phases of the $SU(N)$ links. For these $N$ species, each invariant under a local $U(1)$, magnetic monopoles are identified as in compact QED. The abelian projection additionally yields $SU(N)/[U(1)]^{N-1}$ coset matter fields $c_{ij}(i \neq j)$ corresponding to off-diagonal link matrix elements. Carrying electric charges of species $i$ and $j$, the $c_{ij}$ mediate interspecies interactions.

Species permutation symmetry [5] stipulates that if $\mathcal{O}_i$ denotes an operator comprised exclusively of $i^{\text{th}}$ species abelian links, then $\langle \mathcal{O}_j \rangle = \langle \mathcal{O}_i \rangle$ even if $i \neq j$. As shown in [5] interspecies interactions are $\frac{1}{N}$ suppressed.[1] Hence unless $N \to \infty$ confinement differs fundamentally from finite $N$ confinement interspecies interactions cannot be the chief confinement mechanism.

Therefore, imagining that all else is integrated out let us focus on one representative abelian species. We refer to its field theory as the abelian-projection model of QCD or "APQCD" and its action as $S_{APQCD}$. Numerical studies [4, 6, 7] in lattice $SU(2)$ and $SU(3)$ have shown that APQCD has monopole fluctuations which, analogous to compact QED, are kinetic and dense in the confined phase and static and dilute in the finite temperature phase.

---
[1] While we do not assume "abelian dominance" in this paper, it has been conjectured that the coset fields are unimportant for long-distance physics at all $N$ [2, 6].

Necessary remaining tasks in this program are to demonstrate a causal relation between APQCD monopoles and confinement in the *original* $SU(N)$ gauge theory, to expose the inner workings of this connection, and to understand if and how APQCD confinement survives the continuum limit. To these ends, in this paper we consider a concrete form for $S_{APQCD}$ and determine rough bounds on the parameters of our ansatz.

Since a general lattice action with $U(1)$ local gauge invariance is comprised of arbitrary size and shape Wilson loops in all $U(1)$ representations, and possibly even auxiliary fields, simplifying assumptions must be adopted for progress. Let us suppose $S_{APQCD}$ is comprised only of square plaquettes $P(L) \equiv e^{i\Theta_{P(L)}}$ of size $L \times L$ in lattice units [8]. $\Theta_{P(L)}$ denotes the extended plaquette angle. Neglecting nonlocal interactions we make the ansatz that

$$S_{APQCD} \equiv \sum_{L=1}^{\infty} \sum_{P(L)} s_L(\Theta_{P(L)}). \qquad (1)$$

The sum over $P(L)$ ranges over all $L \times L$ plaquettes in the lattice. By charge conjugation symmetry, $s_L(-\Theta) = s_L(\Theta)$, and gauge invariance [9], $s_L(\Theta + 2\pi) = s_L(\Theta)$, function $s_L$ is Fourier expandable as

$$-s_L(\Theta) \equiv \sum_{q=1}^{\infty} B(q, L) \, \cos(q\Theta) \,. \qquad (2)$$

Section 2 reports on a numerical measurement of the effective 1-plaquette cousin of $B(q, L)$ which, as explained, tends to mimic $B(q, L)$. Our results indicate that $S_{APQCD}$ is dominated by the $B(1,1)$ contribution. Section 3 contemplates implications of our numerical results.

## 2 Effective 1-Plaquette Couplings of APQCD

A typical plaquette $P(L)$ on a $D = 3 + 1$ dimensional lattice shares links with a large number of neighboring plaquettes. Integrating out all links in the lattice except those in $P(L)$ leads to an effective 1-plaquette model. If the gauge group is $U(1)$, its expectation values

$$\langle \mathcal{O}(P(L)) \rangle = \frac{1}{\overline{Z}_L} \int_0^{2\pi} d\alpha \ e^{-\overline{s}_L(\alpha)} \ \mathcal{O}(e^{i\alpha}) \qquad (3)$$

where $\overline{Z}_L = \int_0^{2\pi} d\alpha \ e^{-\overline{s}_L(\alpha)}$ are given by the plaquette spectral density [10]

$$\frac{1}{\overline{Z}_L} e^{-\overline{s}_L(\alpha)} \equiv \overline{\rho}_L(\alpha) \equiv \frac{1}{2\pi} \sum_{\nu=-\infty}^{\infty} e^{-i\nu\alpha} \langle [P(L)]^\nu \rangle. \qquad (4)$$

$\overline{s}_L$ is the 1-plaquette effective action [11]. By analogy[2] to Eq. (2), the 1-plaquette effective coupling constants are defined as

$$\overline{B}(q, L) = \frac{1}{\pi} \int_0^{2\pi} d\alpha \ \log\left(\frac{\overline{\rho}_L(\alpha)}{\overline{\rho}_L(0)}\right) \cos(q\alpha). \qquad (5)$$

If links fluctuate randomly, $\langle [P(L)]^\nu \rangle \to \delta_{\nu,0}, \overline{\rho}_L(\alpha) \to \frac{1}{2\pi}$, and $\overline{B}(q, L) \to 0$. If fluctuations freeze out, then $\langle [P(L)]^\nu \rangle \to 1$, $\overline{\rho}_L(\alpha) \to \sum_{n=-\infty}^{\infty} \delta(\alpha - 2\pi n)$, and $\overline{B}(q, L) \to \infty$. For intermediate cases, suppose momentarily that

$$-S_{APQCD} \sim -S_o \equiv \beta_o \sum_{P(l_o)} \cos(q_o \Theta_{P(l_o)}), \qquad (6)$$

that is, $S_{APQCD}$ is dominated by representation $q_o$ size $l_o \times l_o$ plaquettes. Then at strong coupling($\beta_o \to 0$) effective action $\overline{s}_L$ tends to mimic underlying action $S_o$ as follows. If $L < l_o$, then $\overline{\rho}_L = 0$ since larger plaquettes do not dress smaller ones in the character expansion. If $L \geq l_o$, the planar character expansion yields

$$\overline{\rho}_L(\alpha) = \frac{1}{2\pi} \left(I_0(\beta_o)\right)^{-\left(\frac{L}{l_o}\right)^2} \sum_{\nu=-\infty}^{\infty} e^{-iq_o\nu\alpha} \left(I_\nu(\beta_o)\right)^{\left(\frac{L}{l_o}\right)^2} \qquad (L \geq l_o). \qquad (7)$$

---

[2]To distinguish 1-plaquette quantities from their APQCD cousins we overline the former.

In this approximation effective action $\overline{s}_{l_o} = S_o$ if $L = l_o$ because then $\overline{p}_{l_o}$ resums to $\overline{p}_{l_o}(\alpha) = \frac{1}{2\pi I_0(\beta_o)} e^{\beta_o \cos(q_o \alpha)}$. While Eq. (7) implies that $\overline{B} \neq 0$ even away from $(q_o, l_o)$, "resonance," such operator mixing is significantly suppressed, that is, $\overline{B}(q_o, L \geq l_o) \sim \mathcal{O}(\beta_o^{\left(\frac{L}{l_o}\right)^2})$ and $\overline{B}(q \neq q_o, L) \sim \mathcal{O}(\beta_o^P)$ where typically $P > 10$.

Nonplanar graphs renormalize $\overline{B}$. The one bump correction to the planar graph implies $\overline{B}(q_o, l_o) = \beta_o(1 + \frac{1}{4}\beta_o^4)$.

We have checked these strong coupling arguments against numerical simulations of $D = 3 + 1$ compact QED($q_o = L_o = 1$) over a wide range of $\beta_{QED}$. The intuition gleaned from strong coupling holds qualitatively even in the deconfined phase. Therefore $\overline{B}(q, L)$ resonances extracted from the plaquette spectral densities of APQCD would be evidence of corresponding $B(q, L) \cos(q\Theta_{P(L)})$ terms in $S_{APQCD}$.

To evaluate $\overline{p}_L$ in APQCD, we fix pure gauge $SU(3)$ configurations to MA gauge and perform an abelian projection to get three $U(1)$ fields [4]. For each $U(1)$ field $\langle [P(L)]^\nu \rangle$ and, via Eq. (4), $\overline{p}_L$ are computed. The three $U(1)$ species are averaged for statistical enhancement. Table 1 lists $\overline{B}_{APQCD}(q, L)$ on $\beta_{QCD} = 5.7$, $16^3 \times 24$ and also $\beta_{QCD} = 6.0$, $24^3 \times 40$ lattices [12]. Their compact QED counterparts evaluated on $14^4$ lattices at $\beta_{QED} = .99$(confined region) and $\beta_{QED} = 1.10$(deconfined region) are listed alongside for comparison. While all combinations of $q = 1, \cdots, 20$ and $L = 1, \cdots, 12$ have been evaluated for APQCD, we show only a few values. Note that $\overline{B}(2, 1)$ is positive at $\beta_{QCD} = 5.7$ but negative at $\beta_{QCD} = 6.0$. We believe this is because APQCD does not scale at $\beta_{QCD} = 5.7$. Such scaling violation has been previously detected with other APQCD operators [5].

To verify confinement in APQCD, we plot the APQCD Creutz ratio $\chi_{APQCD}(L, L)$ verses $L$ in Figure 1. For comparison we also perform the abelian projection in Landau gauge. Landau gauge, which does not have residual $U(1) \times U(1)$ gauge symmetry, is not as well motivated as MA gauge.[3] We find that the monopole number density in MA is $1-2$ orders of magnitude larger than in Landau gauge. Correspondingly, MA gauge has greater string tension than Landau gauge in Figure 1. Figure 2 depicts $\|\overline{B}_{APQCD}\|$ on the $\beta_{QCD} = 6.0$ lattice.

| $q$ | $L$ | $\overline{B}_{APQCD}^{\beta_{QCD}=5.7}$ | $\overline{B}_{APQCD}^{\beta_{QCD}=6.0}$ | $\overline{B}_{QED}^{\beta_{QED}=0.99}/\beta_{QED}$ | $\overline{B}_{QED}^{\beta_{QED}=1.10}/\beta_{QED}$ |
|---|---|---|---|---|---|
| 1 | 1 | 2.01(.004) | 3.11(.006) | 1.41(.002) | 2.13(.003) |
| 2 | 1 | .094(.001) | -.057(.003) | -.020(.001) | -0.186(.001) |
| 3 | 1 | -.015(.0005) | .0021(.0020) | .0023(.0006) | .052(.001) |
| 4 | 1 | -.00005(.0006) | .001(.001) | -.0004(.0008) | -.017(.001) |
| 1 | 2 | .669(.003) | 1.40(.003) | .27(.001) | .83(.001) |
| 2 | 2 | -.054(.0006) | -.23(.001) | -.013(.0006) | -.12(.0006) |
| 3 | 2 | .0073(.0003) | .065(.0006) | .0007(.0005) | .029(.0006) |
| 4 | 2 | -.0007(.0004) | -.022(.0003) | -.0001(.0005) | -.0087(.0005) |
| 1 | 3 | .18(.002) | .69(.002) | .034(.0006) | .38(.001) |
| 2 | 3 | -.0069(.0003) | -.090(.0009) | -.0004(.0005) | -.036(.0005) |
| 3 | 3 | .0006(.0003) | .017(.0004) | .00005(.0005) | .0048(.0004) |
| 4 | 3 | -.00005(.0003) | -.0038(.0003) | -.0006(.0004) | -.0002(.0005) |
| 1 | 4 | .036(.0008) | .36(.002) | .0031(.0006) | .19(.001) |
| 2 | 4 | -.0002(.0003) | -.029(.0004) | -.0016(.0003) | -.0096(.0004) |
| 3 | 4 | .00005(.0004) | .003(.0003) | .0005(.0005) | .0006(.0004) |
| 1 | 5 | .0054(.0004) | .19(.002) | -.0002(.0004) | .096(.0008) |
| 2 | 5 | -.00003(.0003) | -.0091(.0002) | .0002(.0005) | -.0028(.0004) |
| 1 | 6 | .0006(.0004) | .10(.001) | -.0005(.0005) | .048(.0006) |
| 2 | 6 | -.0003(.0002) | -.0025(.0002) | -.00009(.0004) | -.00007(.0005) |

Table 1: Couplings $\overline{B}(q,L)$ of 1-plaquette effective actions in $SU(3)$ APQCD and, for comparison, in compact QED. $q$ labels $U(1)$ representations and $L$ is the plaquette width.

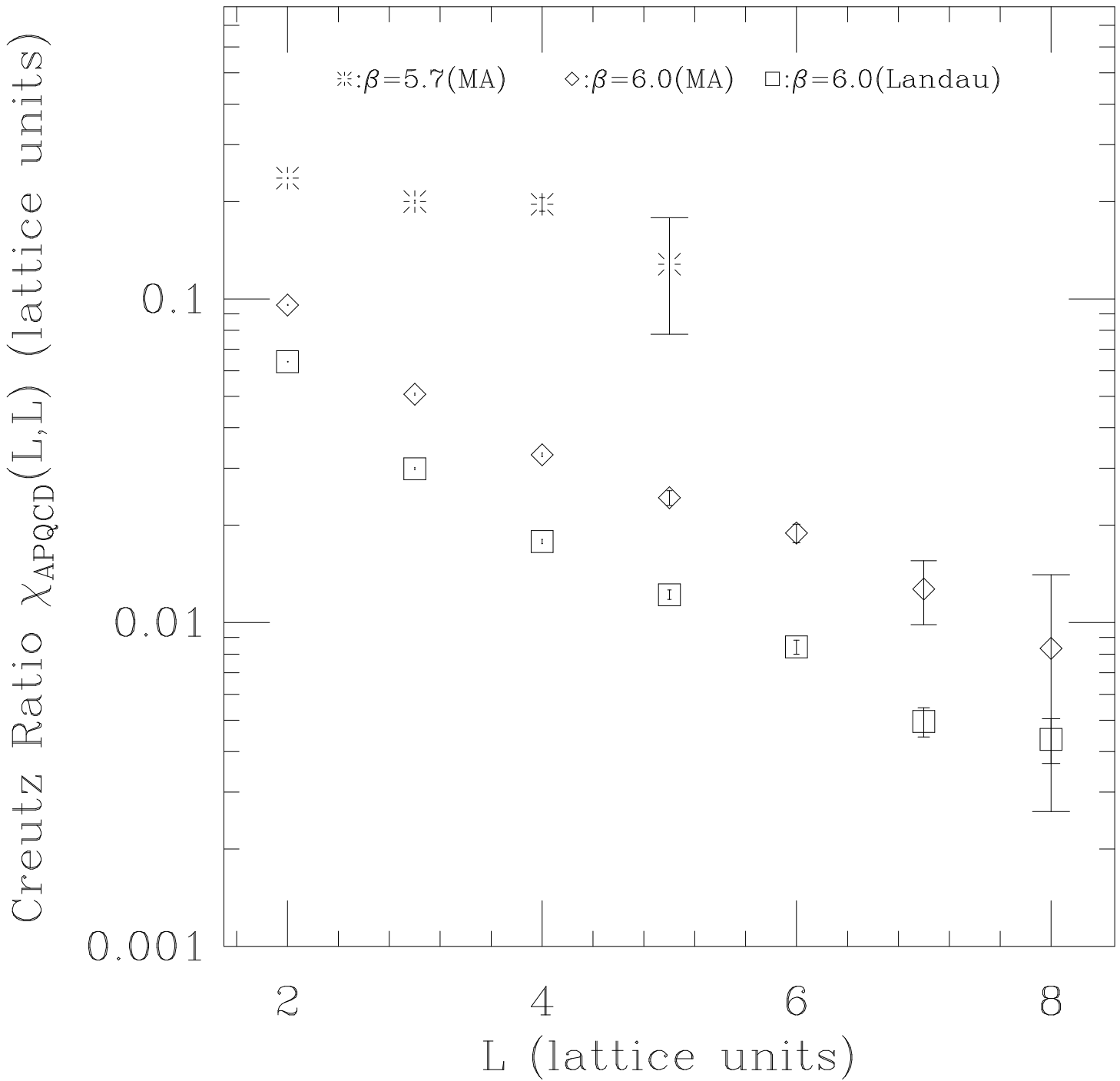

Figure 1: Dimensionless Creutz ratio in APQCD on $\beta_{QCD} = 5.7$, $16^3 \times 24$ and $\beta_{QCD} = 6.0$, $24^3 \times 40$ lattices. Since the ratio of lattice spacings is $a(\beta_{QCD} = 5.7)/a(\beta_{QCD} = 6.0) \sim 2$, the APQCD string tension doesn't seem to scale between these two lattices.

Figure 2: Absolute value $\|\overline{B}_{APQCD}\|$ on the $\beta_{QCD} = 6.0$, $24^3 \times 40$ lattice. Note the jump between $L = 1$ and $L = 2$ on the $q \geq 2$ curves.

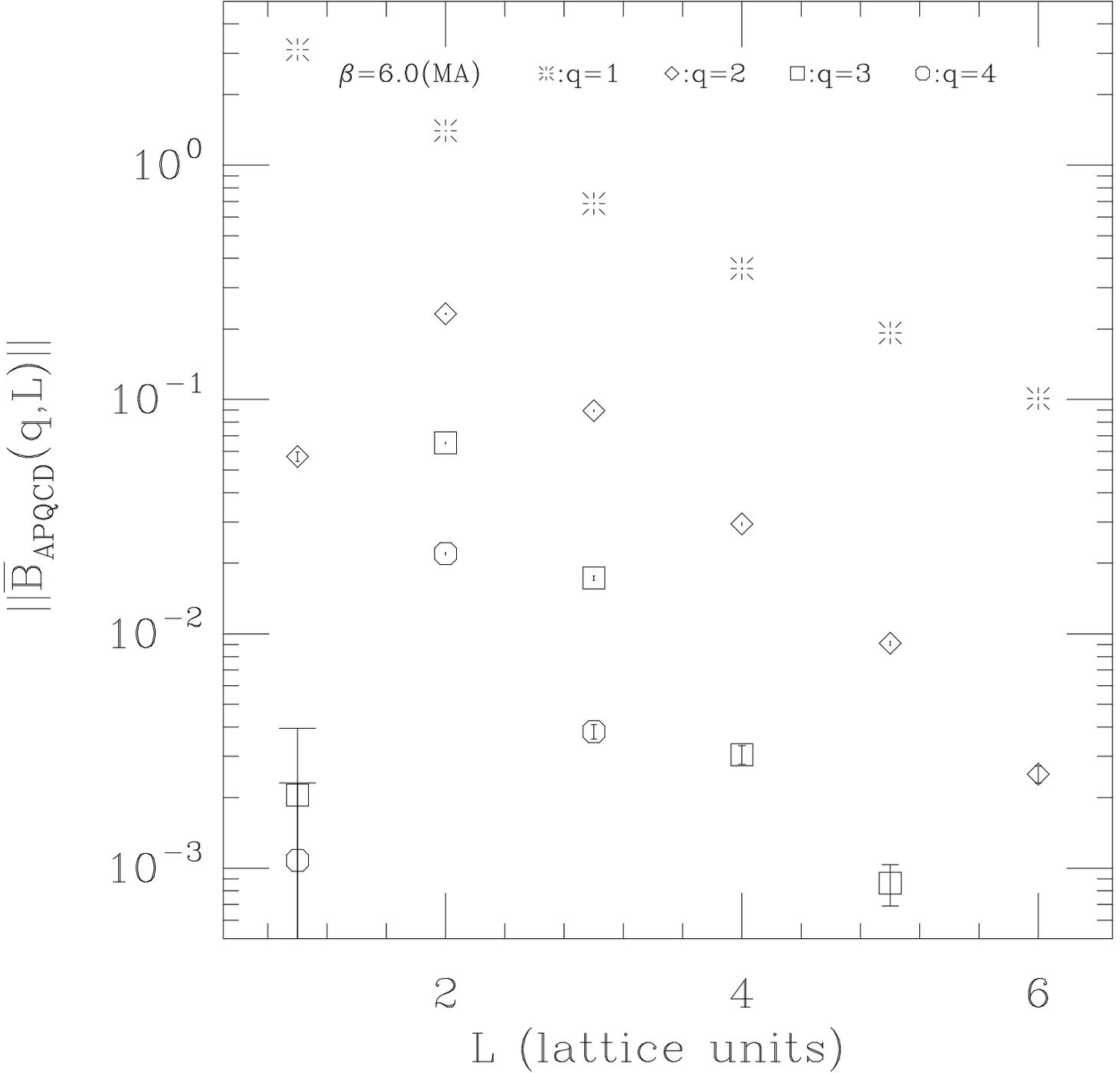

## 3 Discussion

As given in Table 1 and Figure 2 $\|\overline{B}_{APQCD}\|$ is a smoothly decreasing function of $L$ for fixed $q$ and of $q$ for fixed $L$ except that, at $\beta = 6.0$ and $q > 1$, $\|\overline{B}_{APQCD}(q,1)\|$ is less than $\|\overline{B}_{APQCD}(q,2)\|$. It is tempting to surmise that these small resonances at $L = 2$, not present in either phase of compact QED, indicate the presence of weakly coupled $2 \times 2$ plaquettes, either $q = 1$ or $q = 2$ or a combination, in $S_{APQCD}$. In any case, we see no evidence for $L > 1$ or $q > 1$ plaquettes with couplings comparable to $B(1,1)$. Since the abelian projection is a 1-to-1 map of the diagonal phases of fundamental representation $SU(3)$ link matrices into three copies of $U(1)$ and QCD action plaquettes are $L = 1$, it is not surprising that $q = L = 1$ plaquettes dominate APQCD.

What *is* notable is that as given in Table 1

$$\overline{B}_{APQCD}(1,1) > \overline{B}_{QED}^{\beta_{QED}=1.01}(1,1) = 1.86(.010) \qquad (8)$$

and in fact $\overline{B}_{APQCD}^{\beta_{QCD}=6.0}(1,1) \gg \overline{B}_{QED}^{\beta_{QED}=1.10}(1,1)$. Since compact QED does not confine when $\beta_{QED} > 1.01$, the large value of $\overline{B}_{APQCD}(1,1)$ indicates that $S_{APQCD}$ cannot simply be $B_{APQCD}(1,1) \sum_{P(1)} \cos \Theta_{P(1)}$. Action $S_{APQCD}$ *must* contain additional operators for APQCD to be confining.

---

[3]$U(1) \times U(1)$ by itself does not guarantee an equivalent abelian projection [6]. While $U(1) \times U(1)$ symmetry can be restored to Landau gauge configurations with random (Landau gauge violating) $U(1) \times U(1)$ transformations, these transformations cannot alter monopole densities or Creutz ratios, which are $U(1)$ invariant.

What are these additional operators? First, there is the aforementioned evidence for the $2 \times 2$ plaquettes. Second, we have not ruled out the presence of weakly coupled higher $q$ or $L$ plaquettes. It is entirely probable that they are present but that their couplings $B_{APQCD}$ are too small to be easily discerned from effective couplings $\overline{B}_{APQCD}$. Even if this is so, since $D = 3 + 1$ compact QED-like theories in Villain form generically deconfine in the continuum limit [13], it is hard to see how such weakly coupled operators act to nullify the compact QED deconfinement transition as $\beta_{QCD} \to \infty$. In this regard, concrete extensions or counterexamples to Guth's theorem for actions of the form (1) would be useful [14].

Third, as noted, Eq. (1) neglects nonsquare and nonplanar Wilson loops and nonlocal interactions. Our results rule out substantial nonsquare or nonplanar contributions since one expects, for example, that the couplings of nonplanar $3 \times 2$ plaquettes are comparable to planar $3 \times 3$ plaquettes, which are small. On the other hand, our results do not rule out nonlocal interactions such as $\sum_{L,L'} g(L,L') f_L(\Theta_{P(L)}) h_{L'}(\Theta_{P(L')})$ which might arise, for example, due to our hypothetical "integrating out" of the charged $c_{ij}$ coset fields. If this is true, then perhaps APQCD is more naturally formulated as a $[U(1)]^{N-1}$ gauge theory with charged matter fields explicitly present [14]. Of course, since the $U(1)$ gauge fields decouple from each other in the $N \to \infty$ limit, the single $U(1)$ approach must naturalize at large $N$. In this limit, the APQCD view outlined in Section 1 must be adequate.

# 4 Acknowledgments

Analysis was done at NERSC on the author's grant. The author is financially supported by DOE grant DE-FG05-91ER40617.